\documentclass[%
 aip,
 amsmath,amssymb,
 reprint,%
]{revtex4-1}

\usepackage{graphicx}
\usepackage{dcolumn}
\usepackage{bm}

\usepackage[utf8]{inputenc}
\usepackage[T1]{fontenc}
\usepackage{mathptmx}
\usepackage{etoolbox}

\makeatletter
\def\@email#1#2{%
 \endgroup
 \patchcmd{\titleblock@produce}
  {\frontmatter@RRAPformat}
  {\frontmatter@RRAPformat{\produce@RRAP{*#1\href{mailto:#2}{#2}}}\frontmatter@RRAPformat}
  {}{}
}%
\makeatother
\begin{document}

\preprint{AIP/123-QED}

\title{Faster network disruption from layered oscillatory dynamics}
\author{Melvyn Tyloo*}
 \email{mtyloo@lanl.gov.}
\affiliation{ 
$^1$Theoretical Division, Los Alamos National Laboratory, Los Alamos, NM 87545, United States of America \\and Center for Nonlinear Studies (CNLS), Los Alamos National Laboratory, Los Alamos, NM 87545, United States of America
}%

\date{\today}

\begin{abstract}
Nonlinear complex network-coupled systems typically have multiple stable equilibrium states. Following perturbations or due to ambient noise, the system is pushed away from its initial equilibrium and, depending on the direction and the amplitude of the excursion, might undergo a transition to another equilibrium. It was recently demonstrated [M. Tyloo, J. Phys. Complex. {\bf{3}} 03LT01 (2022)], that layered complex networks may exhibit amplified fluctuations. Here I investigate how noise with system-specific correlations impacts the first escape time of nonlinearly coupled oscillators. Interestingly, I show that, not only the strong amplification of the fluctuations is a threat to the good functioning of the network, but also the spatial and temporal correlations of the noise along the lowest-lying eigenmodes of the Laplacian matrix. I analyze first escape times on synthetic networks and compare noise originating from layered dynamics, to uncorrelated noise. 
\end{abstract}

\maketitle

\begin{quotation}
Complex networked systems are often made of different layers of dynamics that somehow interact together. Due to this intricate structure, noise or perturbations affecting one layer are typically transferred to other layers with additional statistical properties e.g. spatial and temporal correlations. The latters might have serious consequences on the desired functioning of some layers, one of them being the amplification of fluctuations.\cite{Tyl22b} Together with the inherent multistability of nonlinearly coupled dynamical systems, noise acting on one layer may be more prompt to drive other layers outside their initial basin of attraction, through a transition to another equilibrium, if such a state exists. Here, I investigate how first escape times are affected by noise propagating through a layered system and compare it to uncorrelated noise.
\end{quotation}

\section{Introduction}
\begin{figure}
    \centering
    \includegraphics[scale=0.4]{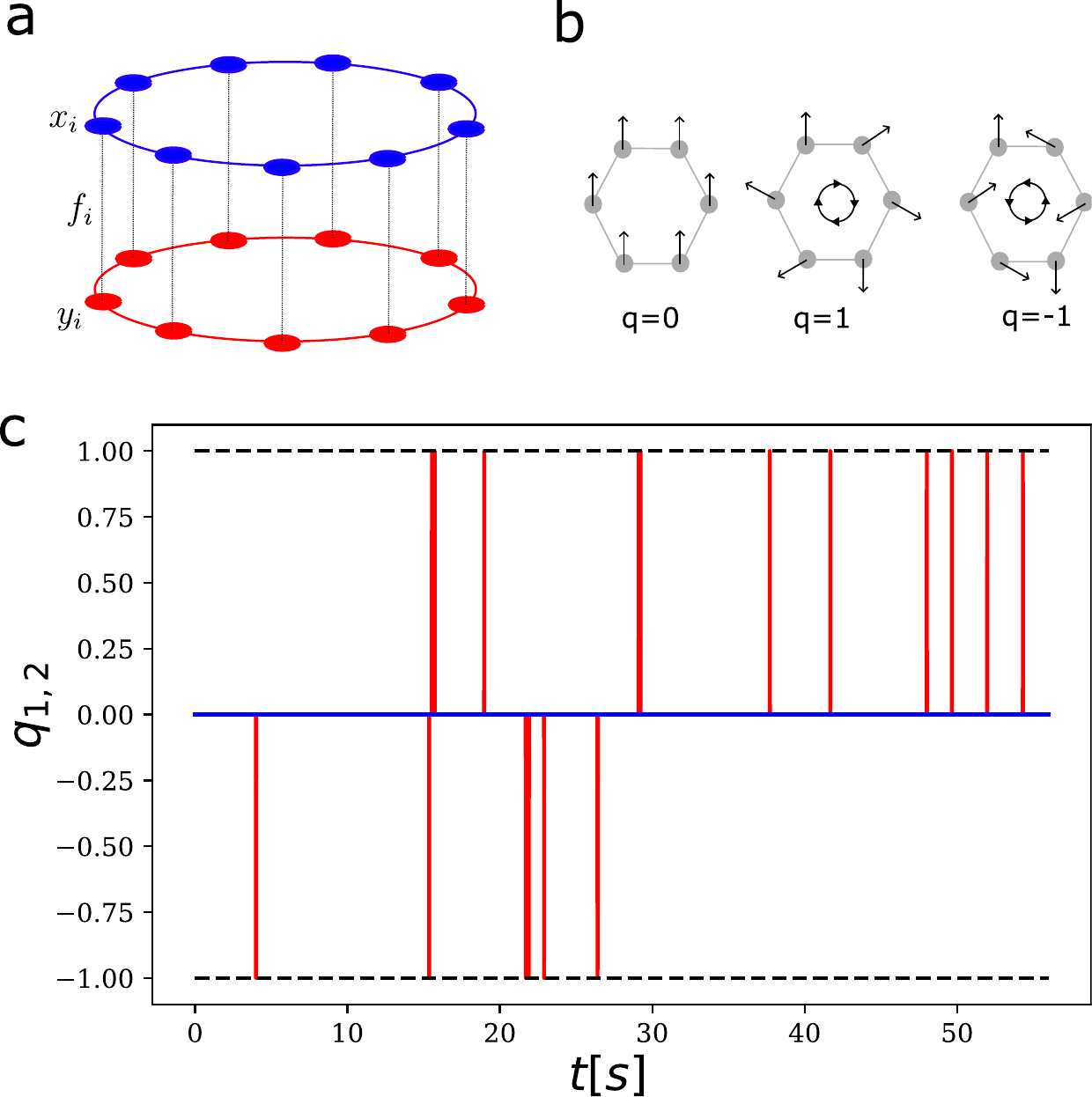}
    \caption{(a) Two-layer system with a single cycle in each layer. (b) Three different fixed points with $q=-1,0,1$ are illustrated for a $n=5$ nodes system, where the orientation of the arrows represents the degrees of freedom. (c) Time-evolution of the winding numbers $q_1$ and $q_2$ in each layer (c) of the two-layers system shown in panel (a), corresponding to Eqs.~(\ref{eq1}) with parameters such that injected noise amplitudes are the same in both layers. The two layers have the same cyclic coupling network of size $n=40$ nodes. Transitions between different basins of attraction are tracked by observing changes in values of $q_{1,2}$\,.  Noise amplitudes are the same in both layers, however their spatial and temporal correlations are different.}
    \label{fig1}
\end{figure}

Various systems in nature and engineered applications can be modelled as individual dynamical units, interacting with one another in a complex way such as neurons spiking together in the brain\cite{Ped18} or rotating masses of power generators evolving at the same frequency in large-scale transmission networks.\cite{Mac08} Collective phenomena taking place in these systems such as synchronization, are enabled by both the internal dynamics and the interaction within the individual units.\cite{Pik03} Another remarkable feature of such systems is multistability. Thanks to nonlinearities in the interaction, numerous equilibria might exist, each of them having their own basin of attraction.\cite{Wil06,Men13} Due to external perturbations such as environmental noise or faults occurring at some components, the whole system might undergo transitions between equilibrium points. In many instances, this is not desirable as both the transient dynamics and the new equilibirum might threaten the correct operation of the system or even cause damages.\cite{Sim08} An important task is to predict such transition as they could disrupt the desired functioning state of the networked system. The latter question has been investigated mostly in coupled systems made of a single layer of interaction, usually subjected to spatially uncorrelated noise.\cite{Dev12,Scha17,Hin18,Tyl18c} However, many applications require more involved coupling structures in order to correctly describe them.\cite{Kur06} A natural extension of the single layer framework, is to include multiple layers of dynamical systems.\cite{BiaB18,Hil20} In such layered dynamics, noise acting on one sub-network propagates to other layers with system-specific correlations.\cite{Tyl22b} In other words, when going through a layer, the noise acquire some specific statistical properties, which is then injected into other layers. Including such system specific correlation in the noise might increase the level of fluctuations and thus the risk of transitions between equilibria. This is depicted in Fig.~\ref{fig1}, where uncorrelated noise is acting on the first layer [blue in panel (a)], which is then transmitted to the second one [red in panel (a)]. In this setting, the second layer is subjected to correlated noise, which we describe below. Both layers are made of a single cycle network which has multiple equilibira as illustrated in Fig.~\ref{fig1}(b). Due to the noise injected in both of them, the two layers are pushed away from their initial equilibrium and, depending on the noise amplitude and direction, eventually leave their initial basin of attraction.  Transitions between basins of attraction can be detected by changes in the winding numbers in each layer $q_1$ and $q_2$ [see Fig.~\ref{fig1}(c)]. One clearly sees that, while the first layer do not operate any transition, the second one exits its initial basin of attraction multiple times. Similar transitions where observed for a single layer system~\cite{Dev12,Hin18} and related to the eigenvalues of the network Laplacian matrix and the distance between the initial fixed point and the closest saddle point with a single unstable direction.\cite{Tyl18c}
Here I consider layered systems where each layer has its own individual units, nonlinearly coupled on a complex network. The different layers interact together via a coupling function. It was recently shown that noise acting on layer might be strongly amplified in the other layers depending on the network connectivities.\cite{Tyl22b} Besides being amplified, the noise structure seems to align with the lowest-lying eigenmodes of the network Laplacian. Building on these results, I investigate the first escape time from the initial basin of attraction in one layer due to noise originating from connected layers. Remarkably, correlations in space and time of this type of noise seem more efficient in driving the system through a transition to another basin than uncorrelated noise. 

The paper is organized as follows. In Sec.~\ref{Sec2}, I introduce the two-layer system of nonlinearly coupled oscillators considered, recalls previous results\cite{Tyl22b} and discuss escape from the basin of attraction. Numerical results are presented in Sec.~\ref{Sec3} and conclusions in Sec.~\ref{Sec4}.

\section{Noisy nonlinear layered oscillators}\label{Sec2}
\subsection{Kuramoto dynamics and its linearization}
In order to investigate the effect of noise in layered networks, I consider a simple two-layer system made of Kuramoto oscillators~\cite{Kur75} as the one shown in Fig.~\ref{fig1}(a). Note that the results below may apply to a larger set of diffusively coupled systems. Their dynamics is governed by $2n$ coupled differential equations, 
\begin{align}\label{eq1}
\begin{split}\dot{x}_i &= \omega_i^{(1)}- \sum_{j=1}^n b_{ij}^{(1)}\,\sin(x_i-x_j) + \eta_i  \quad  i=1,...n\,,\\
\dot{y}_i &= \omega_i^{(2)} - \sum_{j=1}^n b_{ij}^{(2)}\,\sin(y_i-y_j) + f_i(\{x_k\},\{y_k\}) \quad  i=1,...n\,,
\end{split}
\end{align}
where degrees of freedom and natural frequencies respectively in layer 1 and 2 are denoted $\{x_k\}$, $\{y_k\}$ and $\omega_k^{(1)}$, $\omega_k^{(2)}$\,. The undirected coupling network in the $l$-th layer is given by the adjacency matrix elements $b_{ij}^{(l)}\ge 0$, and $f_i$ is a coupling function between the two layers. The noise acting on the first layer is encoded in $\eta_i$ that is taken as white and uncorrelated in space, i.e. $\langle\eta_i(t)\eta_j(t')\rangle = \delta_{ij}\,\eta_0^2\,\delta(t-t')$\,. This model made of Kuramoto oscillators, is a nonlinear version of the one used in Ref.~\citenum{Tyl22b}. Thanks to the sine coupling and provided that $b_{ij}^{(l)}$'s are sufficiently large compared to the distributions of $\omega_i^{(l)}$'s, both Eqs.~(\ref{eq1}) may have multiple stable or unstable fixed points $(\{x_k^{(0)}\},\{y_k^{(0)}\})$\,, depending on the coupling networks. In both cases they satisfy,
\begin{align}\label{eq12}
\begin{split}
0 &= \omega_i^{(1)} - \sum_{j=1}^n b_{ij}^{(1)}\,\sin(x_i^{(0)}-x_j^{(0)})  \quad  i=1,...n\,,\\
0 &= \omega_i^{(2)} - \sum_{j=1}^n b_{ij}^{(2)}\,\sin(y_i^{(0)}-y_j^{(0)}) + f_i(\{x_k^{(0)}\},\{y_k^{(0)}\}) \quad  i=1,...n\,.
\end{split}
\end{align}
In the following, I consider a simple coupling function $f_i(\{x_k\},\{y_k\}) =  d(x_i - n^{-1}\sum_j x_j)$ where degrees of freedom in the first layer tune the natural frequencies in the second one, and $d$ serves as tuning parameter.
The dynamics of small deviations close to a stable fixed point is given by the Taylor expansion of Eqs.~(\ref{eq1}) to the first order in $[\delta x_i(t), \delta y_i(t)]=[x_i(t)-x_i^{(0)}, y_i(t)-y_i^{(0)}]$\,, and read,
\begin{align}\label{eq2}
\begin{split}
\dot{\delta x}_i &= - \sum_{j=1}^n b_{ij}^{(1)}\,\cos(x_i^{(0)}-x_j^{(0)})(\delta x_i-\delta x_j) + \eta_i \\  
\dot{\delta y}_i &= - \sum_{j=1}^n b_{ij}^{(2)}\,\cos(y_i^{(0)}-y_j^{(0)})(\delta y_i-\delta y_j) \\
&\quad+ d(\delta x_i - n^{-1}\sum_j \delta x_j) \,,
\end{split}
\end{align}
which are diffusive linear systems. Indeed, one can define the weighted Laplacian matrices,
\begin{equation*}\label{eq:laplacian1}
{\mathbb L}_{ij}^{(1)}(\{ x_i^{(0)} \}) =
\left\{
\begin{array}{cc}
-b_{ij}^{(1)} \cos(x_i^{(0)} - x_j^{(0)}) \, , & i \ne j \, , \\
\sum_k b_{ik}^{(1)} \cos(x_i^{(0)} - x_k^{(0)}) \, , & i=j \, ,
\end{array}
\right.
\end{equation*}
\begin{equation*}\label{eq:laplacian2}
{\mathbb L}_{ij}^{(2)}(\{ y_i^{(0)} \}) =
\left\{
\begin{array}{cc}
-b_{ij}^{(2)} \cos(y_i^{(0)} - y_j^{(0)}) \, , & i \ne j \, , \\
\sum_k b_{ik}^{(2)} \cos(y_i^{(0)} - y_k^{(0)}) \, , & i=j \, ,
\end{array}
\right.
\end{equation*}
which depend on both the initial coupling networks in each layer and the considered equilibrium.
Using these Laplacian matrices, Eqs.~\ref{eq2} are conveniently rewritten in a vector form as,
\begin{align}\label{eq3}
\begin{split}
\bf{\dot{\delta x}} &= -{\mathbb L}^{(1)}(\{ x_i^{(0)} \})\bf{\delta x} + \bf{\eta} \\  
\bf{\dot{\delta y}} &= - {\mathbb L}^{(2)}(\{ y_i^{(0)} \}){\bf{\delta y}} + {d}\bf{\overline{\delta x}} \,,
\end{split}
\end{align}
where $\overline{\delta x}_i = \delta x_i - n^{-1}\sum_j \delta x_j$\,. From the Laplacian properties of ${\mathbb L}_{ij}^{(1)}(\{ x_i^{(0)} \})$ and ${\mathbb L}_{ij}^{(2)}(\{ y_i^{(0)} \})$, their eigenvalues satisfy: $0=\lambda_1^{(k)}<\lambda_2^{(k)}<...<\lambda_n^{(k)}$ for $k=1,2$ with the first eigenvector being constant i.e. $\bm u_{1}^{(k)}=(1,...,1)/\sqrt{n}$\, for $k=1,2$\,. The latter modes correspond to a symmetry of the system as any shift along it does not modify the dynamics nor the fixed points. The second smallest eigenvalue $\lambda_2^{(k)}$ is usually called \textit{algebraic connectivity} of the network, with its corresponding eigenvector ${\bm u}_2^{(k)}$ called \textit{Fiedler mode}. The larger the algebraic connectivity, the better connected is the network, in the sense that even the slowest modes are quickly damped. The solution of Eqs.~(\ref{eq3}) is obtained by expanding the deviations over the eigenvectors as $\delta x_i = \sum_\alpha c_\alpha^{(1)}u_{\alpha,i}^{(1)}$\,, $\delta y_i = \sum_\alpha c_\alpha^{(2)}u_{\alpha,i}^{(2)}$\,.\\

\subsection{Amplification of the fluctuations}\label{BAf}
\begin{figure}
    \centering
    \includegraphics[scale=0.47]{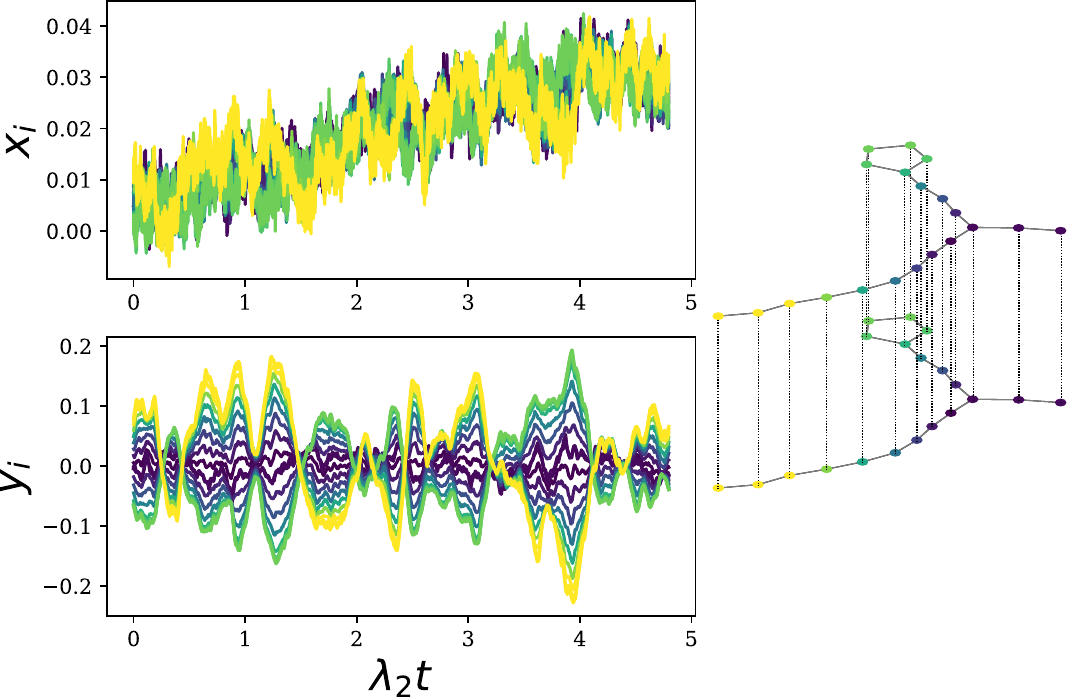}
    \caption{Time-evolution of $x_i$'s and $y_i$'s following Eqs.~(\ref{eq1}) with $d=1$ for the system made of two layers of $n=20$ nodes depicted on the right. The colors represent the squared amplitude of the slowest eigenvector, i.e. $u_{2,i}^2$\, (where I dropped the superscript as the networks are the same in both layers). Lighter colors correspond to larger values of $u_{2,i}^2$ which also have larger standard deviations in the second layer ($y_i$'s).}
    \label{fig2}
\end{figure}

It has been shown recently that, for linear systems governed by the dynamics of Eqs.~(\ref{eq3}), spatially uncorrelated and white in time noise acting on the first layer $x_i$'s might lead to amplified fluctuations in the second layer $y_i$'s.\cite{Tyl22b} Indeed, the variance of the degrees of freedom in the first layer is generated by $\eta_i$'s which are i.i.d. and white in time, and yields\cite{Tyl22b} (see App.~\ref{app1} for details)
\begin{eqnarray}\label{eq4}
\langle \delta x_i^2 \rangle &=& \frac{\eta_0^2}{2}\sum_\alpha \frac{{u_{\alpha,i}^{(1)}}^2}{\lambda_\alpha^{(1)}}\,,
\end{eqnarray}
where $\langle...\rangle$ refers to the average.
In contrast, the variance in the second layer is generated by the $x_i$'s which play the role of additive noise that is both correlated in space and time as shown by their two-point, time shifted correlator,\cite{Tyl22b} 
\begin{eqnarray}\label{eq42}
\langle \delta x_i(t) \delta x_j(t') \rangle = \frac{\eta_0^2}{2}\sum_\alpha \frac{{u_{\alpha,i}^{(1)}}u_{\alpha,j}^{(1)}}{\lambda_\alpha^{(1)}}e^{-\lambda_\alpha^{(1)}|t-t'|}.
\end{eqnarray}
Using the latter correlator together with Eq.~(\ref{eq3}) allows to calculate the variance in the second layer which yields,
\begin{widetext}
\begin{align}\label{eq5}
\begin{split}
\langle \delta y_i^2 \rangle = d^2\,\frac{\eta_0^2}{2}\sum_{\alpha, \beta, \gamma} \sum_{k,l} \frac{{u_{\gamma,k}^{(1)}}u_{\gamma,l}^{(1)} {u_{\alpha,k}^{(2)}}u_{\beta,l}^{(2)} [2\lambda_\gamma^{(1)} +\lambda_\alpha^{(2)} + \lambda_\beta^{(2)}] }{\lambda_\gamma^{(1)} (\lambda_\alpha^{(2)} + \lambda_\beta^{(2)} ) ( \lambda_\gamma^{(1)} + \lambda_\alpha^{(2)} ) ( \lambda_\gamma^{(1)} + \lambda_\beta^{(2)} )  }{u_{\alpha,i}^{(2)}}u_{\beta,i}^{(2)} \,.
\end{split}
\end{align}
\end{widetext}
For linearly coupled oscillators, the variance is exactly described by Eqs.~(\ref{eq4}), (\ref{eq5}). Depending on the overlap between the slowest eigenvectors of ${\mathbb L}_{ij}^{(1)}(\{ x_i^{(0)} \})$ and ${\mathbb L}_{ij}^{(2)}(\{ y_i^{(0)} \})$\,, fluctuations in the second layer might be strongly amplified. In the particular case where the networks are same in both layers, i.e. ${\mathbb L}_{ij}^{(1)}(\{ x_i^{(0)} \})={\mathbb L}_{ij}^{(2)}(\{ y_i^{(0)} \})$\,, Eq.~(\ref{eq5}) simplifies to,
\begin{align}\label{eq6}
\begin{split}
\langle \delta y_i^2 \rangle = d^2\,\frac{\eta_0^2}{4}\sum_{\alpha}  \frac{u_{\alpha,i}^2}{\lambda_\alpha^3} \,.
\end{split}
\end{align}
Due to the third power of $\lambda_{\alpha}$ at the denominator, the variance is dominated by the slowest modes. So not only the overall amplitude of the fluctuations might be large, but also their shape follows those of the slowest modes. This is shown in Fig.~\ref{fig2}. For both $x_i$'s and $y_i$'s, an important contribution to the variance comes from $u_{2,i}^2$ [see Eqs.~(\ref{eq4}), (\ref{eq6})]. However, for the second one this contribution dominates even more due to the denominator in Eq.~(\ref{eq6}). One clearly sees that nodes corresponding to the largest $u_{2,i}^2$'s (lighter colors) have the largest standard deviations in the second layer. Therefore, the second layer is perturbed away from its initial equilibrium mostly along the directions of the slowest eigenmodes.
\subsection{Escape from the basin of attraction}\label{escape}
While the linear system on the first line of Eq.~(\ref{eq3}) have a single fixed point (neglecting the symmetry along ${\bm u}_1^{(1)}$), it is not the case of nonlinearly coupled systems which may have multiple stable fixed points each of them with their own basin of attraction. If the excursion from the initial equilibrium becomes too important, the system undergoes a transition to another basin of attraction corresponding to another fixed point or reach an unstable region of the phase space. Obviously, large perturbations leading to important excursions result in short first escape times. However, not only the amplitude but also the direction of the perturbation matter. Indeed, basins of attraction are not isotropic\cite{Wil06,Del17b,ZhangStro21} and excursion along specific directions may lead to faster transitions. This is already pointed at by Fig.~\ref{fig1}(c) where the amplitudes of the noise acting on layer 1 and 2 are the same. However, the second layer (in red) operates more transitions than the first one (in blue) which stays in its initial basin of attraction. 

In order to detect escapes, one can use the winding number defined on a cycle $c$ as,
\begin{eqnarray}
q = (2\pi)^{-1}\sum_{i\in c}|z_i-z_{i+1}|_{[-\pi, \pi)}\,,
\end{eqnarray}
where $|.|_{[-\pi, \pi)}$ brings the argument into the interval $[-\pi, \pi)$ and $z_i$ is the degree of freedom of the $i$-th oscillator in the cycle. The winding number is therefore an integer given by the number of times that the degrees of freedom wind around the origin along a cycle $c$ [see Fig.~\ref{fig1}(b)]. On a planar meshed network, one can further define a winding vector $\bf{q}$ whose components are given by the winding number on each cycle composing the network.  
Each stable fixed point of Eqs.~(\ref{eq1}) are unambiguously identified by a pair of winding vectors $({\bf{q_1}}, {\bf{q_2}})$ where $\bf{q_l}$ is the collection of winding numbers on each cycle of the $l$-th layer. Therefore, by recording $({\bf{q_1}}, {\bf{q_2}})$\,, one is able to detect basin's escapes independently in each layer. This technique is useful in cases where one is able to measure the flows on each edge, and therefore, determine the winding numbers on each cycle. Another way of measuring first escape times is simply by letting the system relax at every time step and checking that the stable fixed point is the same as the original. This is for sure more precise in determining the first escape time, however, it is much more costly in terms of numerical simulations.

\section{Numerical simulations}\label{Sec3}
In this section, I compare the first escape time in both layers varying the settings. I first consider a system made of two times the same network. To disentangle the effects of amplitude and shape of the noise coming from the first layer, I investigate the situations where (i) the noise is not rescaled, i.e.  large fluctuation might appear in the second layer; (ii) rescaled noise in the second layer so that noise amplitudes in both layers are the same on average, but their shapes are different. The latter is done by tuning the $d$ parameter in the inter-layer coupling function. More details are given below. Then I briefly consider the case of different networks. The method to detect escape based on winding numbers is used for cycle networks, and a combination or the more computationally expensive one and the winding numbers for more complex networks (see Sec.~\ref{escape}). For simplicity, I consider identical oscillators i.e. $\omega_i^{(l)}=0$\,, $\forall i,l$\,. 
\begin{figure}
    \centering
    \includegraphics[scale=0.53]{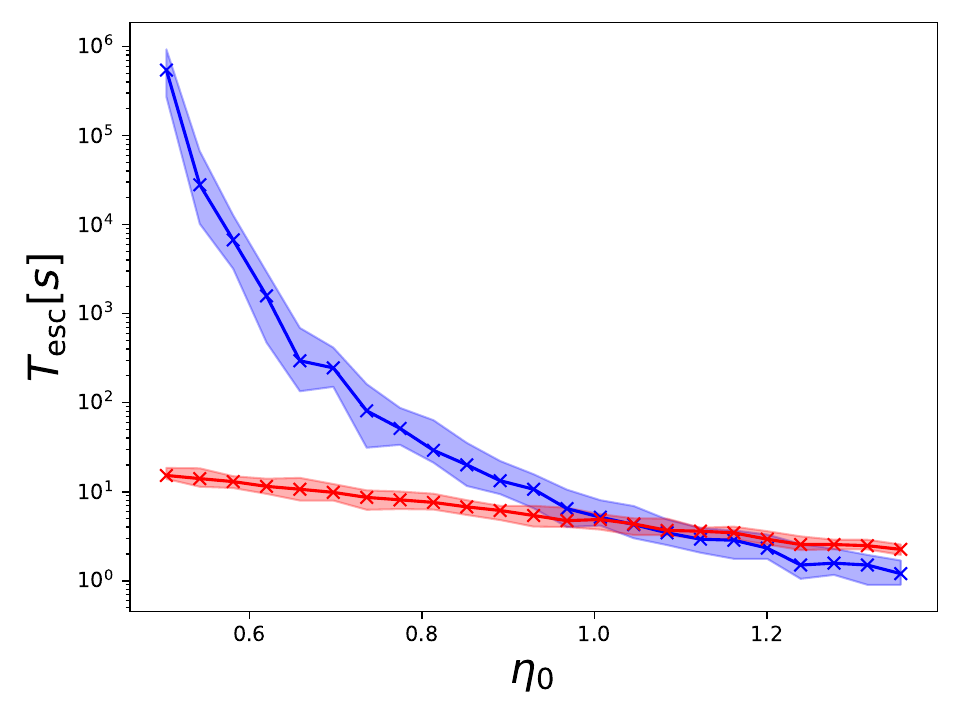}
    \caption{First escape time $T_{\rm{esc}}$ as a function of the noise amplitude in the first layer $\eta_0$\, with $d=1$\,. This system is composed of two cycle networks of size $n=40$ in each layer. Each cross gives the median over 40 realizations. The solid lines show the median first escape time for the first (blue) and second (red) layer. Shaded areas show the 0.25- and 0.75-quantiles.}
    \label{fig3}
\end{figure}
\subsection{Same networks}
When the two networks are the same, the noise injected in the second layer has statistical properties that are directly connected to its structure [see Eq.~(\ref{eq42})]. One therefore expects an important difference in the first escape times of the two layers. More precisely, for networks with low algebraic connectivity, fluctuations are amplified in the second layer and therefore, one expects shorter first escape times for the second layer compared to the first one. For network with high algebraic connectivity, one expects the opposite, as fluctuations are reduced in the second layer.
\subsubsection{Amplification}\label{secn1}
Here we set $d=1$ in the coupling function, meaning that the noise in the second layer has both a specific modal structure and is potentially strongly amplified or reduced. In this case, the two networks are the same and, therefore, share the same eigenvectors and eigenvalues. One expects enhances fluctuations in the second layer if the algebraic connectivity is less than one i.e. $\lambda_2\ll 1$\,.\cite{Tyl22b} This is in particular the case for large cycle networks whose algebraic connectivity scales with $n$ as $\lambda_2 = 2-2\cos(2\pi/n)$\,. Here I focus on such networks with amplification, as those with $\lambda_2\gg 1$ have reduced fluctuations in the second layer and thus, are less likely to leave the initial basin of attraction. Fig.~\ref{fig3} gives the first escape times for both layers in a system made of two cycles. One clearly sees that the first escape time is shorter by several orders of magnitudes in the second layer compared to the first one for noise amplitude $\eta_0<1.1$\,. Then for larger $\eta_0$ the first layer exits its initial basin of attraction faster than the second one. In this regime, the dynamics in the first layer is essentially dominated by the noise with only little influence coming from the coupling network. The noise injected in the second layer may therefore not have any amplification nor specific structure in this case as the linearized calculation presented here does not apply anymore. 

\subsubsection{Rescaled noise}\label{secn2}
\begin{figure}
    \centering
    \includegraphics[scale=0.53]{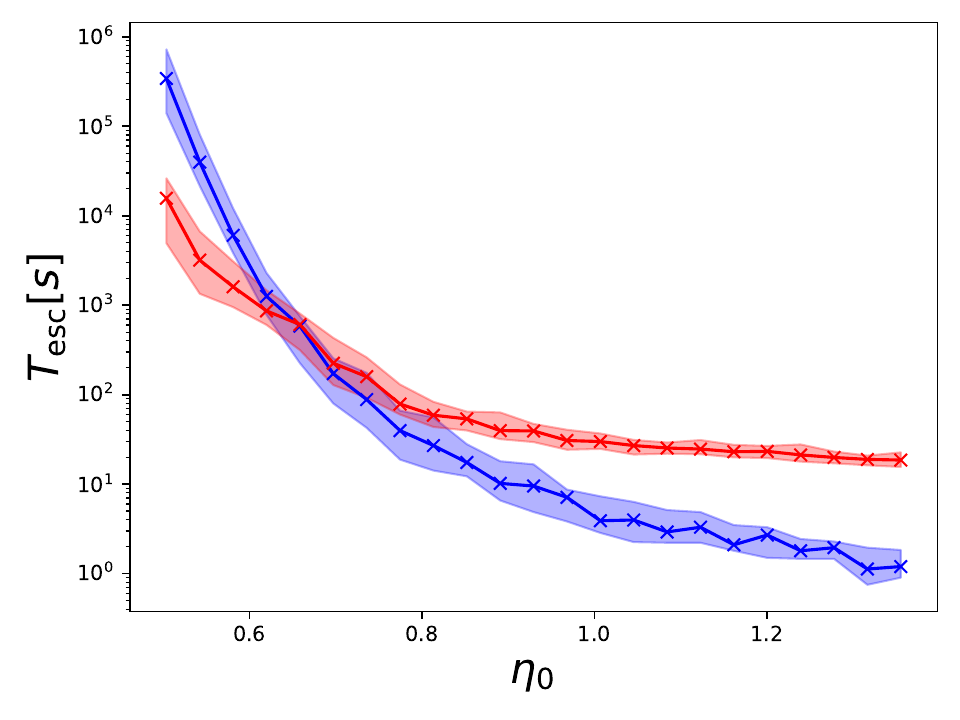}
    \caption{First escape time $T_{\rm{esc}}$ as a function of the noise amplitude in the first layer $\eta_0$\, with $d=\left(2/\sum_\alpha {\lambda_\alpha^{(1)}}^{-1}\right)^{1/2}$\,. This system is composed of two cycle networks of size $n=40$ in each layer. Each cross gives the median over 40 realizations. The solid lines show the median first escape time for the first (blue) and second (red) layer. Shaded areas show the 0.25- and 0.75-quantiles.}
    \label{fig4}
\end{figure}

In order to differentiate the effect of noise amplitude and shape, one needs to rescale the noise coming from first layer so that the amplitude of the noise in the second layer is smaller.
The noise coming from the first layer and to which is subjected the second layer is given by ${\bf{\xi}}={d}\,\bf{\overline{\delta x}}$\,. The noise variance in the first layer is $\eta_0^2$\,. Choosing $d^2=2/\sum_\alpha {\lambda_\alpha^{(1)}}^{-1}$ produces an input noise in the second layer that has on average a variance of $\eta_0^2$\,, i.e.
\begin{eqnarray}\label{resc}
n^{-1}\sum_i \langle \xi_i^2 \rangle = \eta_0^2/n\,,
\end{eqnarray}
so that the second layer is subjected to a noise of variance on average $n$ times smaller than the first layer and have different shape and time correaltions [see Eq.~(\ref{eq42})]. This is illustrated in Fig.~\ref{fig4} for a system made of two cycles, where the noise in the second layer is rescaled. Interestingly, one observes that the first escape time in the second layer is still shorter than in the first one for $\eta_0<0.65$\,. This means that the structure of noise as discussed in Sec.~\ref{BAf} favors faster escapes from the initial basin of attraction compared to uncorrelated noise.

\subsection{Different networks}
For two different networks, noise transmission essentially depends on the overlap between $u_{\alpha,l}^{(1)}$ and $u_{\beta,l}^{(2)}$ for the slowest modes. If the overlap is low, one expects that the noise is transmitted mostly without amplification and therefore, first escape times should be similar in both layers. In the other situation where the overlap is significant, one expects different first escape times in the two layers.

\subsubsection{Amplification}
\begin{figure}
    \centering
    \includegraphics[scale=0.53]{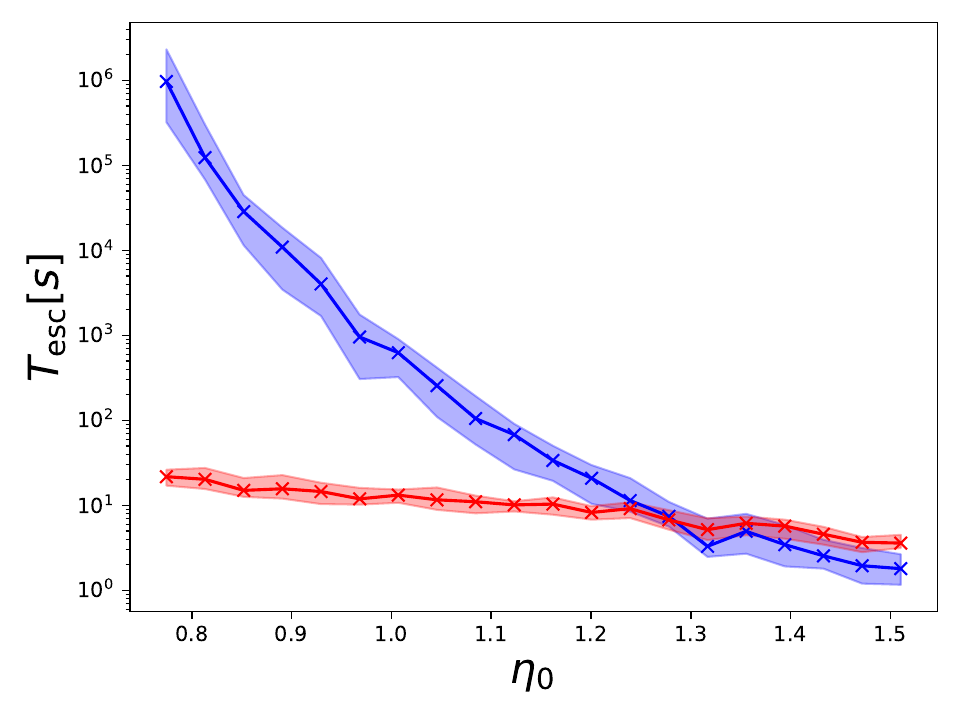}
    \caption{First escape time $T_{\rm{esc}}$ as a function of the noise amplitude in the first layer $\eta_0$\, with $d=1$\,. This system is composed of two different networks of size $n=80$ in each layer, which have been generated using Watts-Strogatz procedure starting from coupling to first and second nearest neighbors and rewiring probabilities $p=0.01$ (first layer), $p=0.04$ (second layer). Each cross gives the median over 40 realizations. The solid lines show the median first escape time for the first (blue) and second (red) layer. Shaded areas show the 0.25- and 0.75-quantiles.}
    \label{fig5}
\end{figure}

Similar effects as in Sec.~\ref{secn1} can be observed when networks are different in the two layers. Here, I consider two networks generated following Watts-Strogatz~\cite{Wat98} procedure with first and second nearest neighbor couplings and low rewiring probabilities. Doing so, the algebraic connectivities of both networks are smaller than one, i.e. $\lambda_2^{(1)}= 0.033$ $\lambda_2^{(2)}= 0.084$\,, and the overlap between their slowest eigenvectors is sufficiently different from zero i.e. $\sum_i u_{2,i}^{(1)}u_{2,i}^{(2)}=0.4$\,.  Therefore, fluctuations are expected to be amplified in the second layer. Indeed, Fig.~\ref{fig5} shows the first escape time in both layers, and the second layer leaves the initial basin of attraction many orders of magnitudes faster than the first one for low values of $\eta_0$\,.

\subsubsection{Rescaled noise}
\begin{figure}
    \centering
    \includegraphics[scale=0.53]{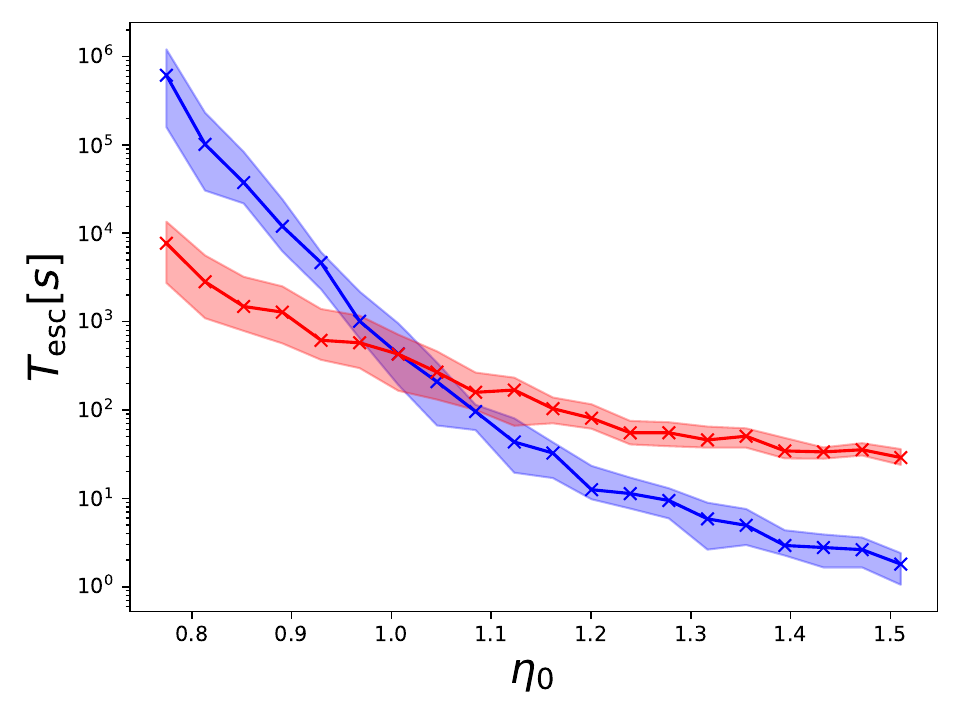}
    \caption{First escape time $T_{\rm{esc}}$ as a function of the noise amplitude in the first layer $\eta_0$\, with $d=\left(2/\sum_\alpha {\lambda_\alpha^{(1)}}^{-1}\right)^{1/2}$\,. This system is composed of two different networks of size $n=80$ in each layer, which have been generated using Watts-Strogatz procedure starting from coupling to first and second nearest neighbors (same networks as Fig.~\ref{fig5}). Each cross gives the median over 40 realizations. The solid lines show the median first escape time for the first (blue) and second (red) layer. Shaded areas show the 0.25- and 0.75-quantiles.}
    \label{fig6}
\end{figure}

Following the same steps as in Sec.~\ref{secn2} to emphasize the effect of the shape of the noise, I rescale the noise such that Eq.~(\ref{resc}) is satisfied. This allows to have injected noise in the second layer that has on average a variance $n$ times smaller than the first one. Remarkably, Fig.~\ref{fig6} shows that the first escape time in the second layer is still shorter than in the first layer, despite the rescaling of the noise for $\eta_0<1$\,. The shape that the noise acquire when passing through the first layer seems to be more efficient in driving the second layer outside its initial basin of attraction, as in the previous case of two times the same network.

\section{Conclusion}\label{Sec4}
Noisy layered systems can exhibit amplified fluctuation patterns depending on their connectivity. Here I showed that noise originally injected in one layer may induce faster basin escape in connected layers. This is both due to the amplification of the noise amplitude and the system specific correlations that the noise acquire while going through the first layer. Indeed, from Eq.~(\ref{eq42}) one sees that the noise in the second layer is correlated in both space and time with clear dependence on the network structure. For networks with low algebraic connectivity, I numerically showed that the first escape time is shorter in the two cases where (i) fluctuations are amplified in the second layer and (ii) noise in the second layer is rescaled in order to have a variance $n$ times smaller than the first layer.
While point (i) is rather intuitive, i.e. larger fluctuations lead to shorter first escape times, point (ii) is more involved. Indeed, this indicates that noise with spatial and temporal correlations given by Eq.~(\ref{eq42}) selects directions that enable faster exits from the initial basin of attraction. 

Even though I consider Kuramoto oscillators, the results presented here may apply to other nonlinear coupling functions as long as the system has an equilibrium point. Further studies should consider time-correlated noise injected in the first layer, and the effect of inertia.\\

\section*{Acknowledgements}
This work was supported by U.S. DOE/OE as part of the DOE Advanced Sensor and Data Analytics Program and by the Laboratory Directed Research and Development program of Los Alamos National Laboratory under project numbers 20220797PRD2 and 20220774ER.

\renewcommand{\theequation}{A.\arabic{equation}}

\setcounter{equation}{0}
\section*{Appendix A: Calculation details to obtain Eq.~(\ref{eq4})}\label{app1}
In order to calculate $\langle \delta x_i^2 \rangle$\,, one needs the solution to the first line of Eq.~(\ref{eq3}). The latter is obtained by expanding the $\delta x_i$'s over the eigenvectors of $\mathbb{L}^{(1)}$\,, i.e. $\delta x_i = \sum_\alpha c_\alpha^{(1)}u_{\alpha,i}^{(1)}$\,, which yields the differential equations,
\begin{eqnarray}
\dot{c}_\alpha^{(1)} + \lambda_\alpha^{(1)}\, c_\alpha^{(1)} = \sum_i \eta_i\, u_{\alpha,i}^{(1)}\,, \quad \alpha=1,...,n\,.
\end{eqnarray}
The solution of the latter equations is given by,
\begin{eqnarray}
{c}_\alpha^{(1)}(t) = e^{-\lambda_\alpha^{(1)}t}\int_0^t e^{\lambda_\alpha^{(1)}t'} \sum_i \eta_i\, u_{\alpha,i}^{(1)}\,\rm{d}t'\,,
\end{eqnarray}
where the initial condition is $c_\alpha^{(1)}(t=0)=0\,$ $\forall \alpha$\,. Then, the long time limit of the variance in the first layer is given by,
\begin{eqnarray}
\langle \delta x_i^2 \rangle &=& \sum_{\alpha,\beta} \langle c_\alpha^{(1)}c_\beta^{(1)}\rangle u_{\alpha,i}^{(1)}u_{\beta,i}^{(1)}\\
&=& \sum_\alpha e^{-2\lambda_\alpha^{(1)}t}\int_0^t e^{2\lambda_\alpha^{(1)}t'}\eta_0^2\, {u_{\alpha,i}^{(1)}}^2\,\rm{d}t'\,,
\end{eqnarray}
where in the last equality I used $\langle\eta_i(t)\eta_j(t')\rangle = \delta_{ij}\,\eta_0^2\,\delta(t-t')$ and the orthogonality between the eigenvectors. Performing the integral and taking the limit of long time gives the results of Eq.~(\ref{eq4})\,.



%


\end{document}